\documentclass[aps, prb, twocolumn, floatfix, amssymb]{revtex4}

\usepackage{graphicx}

\begin{document}

\title{Bolometric technique for high-resolution broadband microwave spectroscopy of ultra-low-loss samples}

\author{P. J. Turner}
\author{D. M. Broun}
\altaffiliation[Permanent address: ]{Dept. of Physics, Simon Fraser University,
Burnaby, BC, Canada, V5A 1S6.}
\author{Saeid Kamal}
\altaffiliation[Permanent address: ]{Dept. of Physics, Simon Fraser University,
Burnaby, BC, Canada, V5A 1S6.}
\author{M. E. Hayden}
\altaffiliation[Permanent address: ]{Dept. of Physics, Simon Fraser University,
Burnaby, BC, Canada, V5A 1S6.}
\author{J. S. Bobowski}
\author{R. Harris}
\author{D. C. Morgan}
\author{J. S. Preston}
\altaffiliation[Permanent address: ]{Dept. of Physics and Astronomy, McMaster
University, Hamilton, ON, Canada, L8S 4M1.}
\author{D. A. Bonn}
\author{W. N. Hardy}
\affiliation{Department of Physics and Astronomy, University of British
Columbia, Vancouver, B.C., Canada  V6T 1Z1}
\date{\today}

\begin{abstract}

A novel low temperature bolometric method has been devised and implemented for
high-precision measurements of the microwave surface resistance of small
single-crystal platelet samples having very low absorption, as a continuous
function of frequency. The key to the success of this non-resonant method is
the \textit{in-situ} use of a normal metal reference sample that calibrates
the absolute rf field strength. The sample temperature can be controlled
independently of the 1.2~K liquid helium bath, allowing for measurements of the
temperature evolution of the absorption. However, the instrument's sensitivity decreases at higher temperatures, placing a limit on the useful temperature range. Using this method, the minimum detectable power at 1.3~K is 1.5~pW,
corresponding to a surface resistance sensitivity of $\approx$1~$\mu\Omega$ for
a typical 1~mm$\times$1~mm platelet sample.

\end{abstract}

\maketitle

\section{Introduction}

In this article we describe an apparatus designed for the continuous-frequency
measurement of low temperature electromagnetic absorption spectra in the
microwave range. The motivation to develop this instrument comes from a desire
to resolve, in great detail, the microwave conductivity of high-quality
single crystals of high-T$_c$ cuprate superconductors.  However, the technique we describe should find a wealth of applications to other condensed
matter systems, providing a means to explore the dynamics of novel electronic states with unprecedented resolution. The possibilities include: the physics
of the metal--insulator transition, where charge localization should lead to frequency
scaling of the conductivity; electron spin resonance spectroscopy of crystal
field excitations; ferromagnetic resonance in novel magnetic structures; and
cyclotron resonance studies of Fermi surface topology.  In addition to the cuprate superconductors, other natural possibilities in the area of superconductivity include heavy fermion and ruthenate materials, as well as high resolution spectroscopy of low-frequency collective excitations such as  Josephson plasmons and order-parameter collective modes.

Early microwave measurements on high quality single crystals of
YBa$_2$Cu$_3$O$_{7-\delta}$ showed that
cooling through T$_c\approx 90$~K  decreased the surface resistance very rapidly, by four orders of magnitude at 2.95~GHz, reaching a low
temperature value of several $\mu\Omega$.\cite{bonn} Resolving
this low absorption in the microwave region has provided a technical challenge
that has been successfully met over most of the temperature range below T$_c$
by the use of high precision cavity-perturbation techniques.\cite{ginsberg, ormeno} In these
experiments, the sample under test is brought into the microwave fields of a
high quality-factor resonant structure made from superconducting cavities or low-loss
dielectric pucks. A limitation of such techniques is
that the resonator is generally restricted to operation at a single fixed frequency,
therefore requiring the use of many separate experiments in order to reveal a spectrum. Furthermore, a very general limitation of the cavity perturbation method is that the dissipation of the unknown sample must exceed the dissipation of the cavity itself in order to be measured with high precision --- a very strong demand for a high quality superconductor in the $T\to0$ limit. The measurement of the residual absorption in superconductors is challenging at any frequency: in the case of infra-red spectroscopy the problem becomes that of measuring values of reflectance that are very close to unity. The challenge lies in the calibration of the measurement, and in both microwave and infra-red work, one relies on having a reference sample of known absorption to calibrate the loss in the walls of the microwave resonator or the infra-red reflectance. Despite these limitations, resonant microwave techniques are the only methods with sufficient sensitivity to measure the evolution of the microwave absorption over a wide temperature range.  In a recent effort by our group, five superconducting resonators were used to map a coarse conductivity spectrum from 1~GHz to 75~GHz in
exceptionally clean samples of YBa$_2$Cu$_3$O$_{6.99}$ from 4~K to 100~K.\cite{hosseini}
This work revealed low temperature quasiparticle dynamics inconsistent with simple models
of $d$-wave superconductivity, whose key signatures appear in the frequency
dependence of the conductivity.\cite{atkinson,hettler,berlinsky1}  The failure of simple theories to give a complete description of the temperature evolution and shape of the conductivity spectra in the best quality samples has driven us to develop the technique described here, with the result that we can now resolve low-temperature microwave conductivity spectra in unprecedented detail.

Bolometric detection is a natural method for measuring the surface resistance spectrum over a continuous frequency range.  For any conductor, the power absorption in a microwave magnetic field is directly proportional to the surface
resistance $R_s$:
\begin{equation}
\label{eqn:power} \ P_{abs}= R_s \int H_{rf}^2 \textrm{d}S,
\end{equation}
where $H_{rf}$ is the r.m.s. magnitude of the tangential magnetic field at the surface $S$. As a
result, a measurement of the temperature rise experienced by a weakly-thermally-anchored
sample exposed to a known microwave magnetic field $H_{rf}$ directly gives $R_s$. To enhance rejection of spurious temperature variations, the rf power should be amplitude
modulated at low frequency and the resulting temperature
oscillations of the sample detected synchronously.

We note that as part of a pioneering study of superconducting Al, a similar bolometric microwave
technique was used by Biondi and Garfunkel to examine the detailed temperature dependence
of the superconducting gap frequency.\cite{BiondiandGarfunkel} This earlier experiment had
the simplifying advantage of measuring the absorption by a large waveguide made
entirely from single crystalline Al. Unfortunately, in more complicated
materials such as the multi-elemental cuprate superconductors, the best quality
samples can only be produced as small single crystals. More recently,
frequency-scanned bolometric measurements have proven useful in probing
collective excitations in small samples of high-T$_c$ cuprates at
frequencies above 20~GHz where the absorption is larger and much easier to
measure.\cite{ophelia} These techniques, however, have not focussed on the
challenge of resolving the low temperature absorption of high-quality single
crystals across a broad frequency range.

A characteristic feature of many electronic materials of current interest is
reduced dimensionality, which gives rise to highly anisotropic transport coefficients.
When making microwave measurements, a well-defined geometry must be chosen in order to separate the individual components of the conductivity tensor, and also to ensure that demagnetization effects do not obscure the measurement. One particularly
clean approach that has been widely used places the sample to be characterized near a position of high symmetry in a microwave enclosure, in the quasi-homogeneous microwave magnetic field near an electric node. Often, single crystal samples grow naturally as platelets having a broad $\hat{a}$--$\hat{b}$ plane crystal face and thin $\hat{c}$-axis dimension, and demagnetization effects are minimized if the broad face of the sample is aligned parallel to the field. In response to the applied rf magnetic field, screening currents flow near the surface of the sample along the broad $\hat{a}$ or $\hat{b}$ face and must necessarily flow along the $\hat{c}$ direction to complete a closed path. In some cases, it is desirable to work with samples that are very thin, rendering the $\hat{c}$-axis contribution negligible. Alternatively, by varying the aspect ratio of the sample by either cleaving or polishing, one can make a series of measurements to disentangle the different crystallographic contributions, without having to change samples. For example, in the cuprate superconductors, the conductivity parallel to the  two dimensional CuO$_2$ plane layers can be several orders of magnitude larger than that perpendicular to the weakly-coupled planes. Typical as-grown
crystal dimensions are $1.0\times1.0\times.01$~mm$^3$. For this aspect
ratio, experiments where a sample was cleaved into many pieces showed that the $\hat{c}$-axis contribution is unimportant.\cite{caxis} We note here that all measurements presented in this article employ the low-demagnetization sample orientation discussed above.

The broadband surface-resistance measurement technique we describe in the
following sections provides three distinct technical advances over previous
bolometric approaches: a uniform microwave field configuration in the sample
region that permits the separation of anisotropic conductivity components;
the use of an \textit{in-situ} reference sample that calibrates the
microwave field strength at the sample absolutely; and very high sensitivity
afforded by the choice of a resistive bolometer optimized for the low-temperature range and mounted on a miniaturized thermal stage. These features of our apparatus permit precision measurements of the absolute value of $R_s(\omega,T)$ in very-low-loss samples down to 1.2~K and
over the frequency range 0.1-21~GHz. We will briefly demonstrate that this
range captures the key frequency window for long-lived nodal quasiparticles in
extremely clean samples of YBa$_2$Cu$_3$O$_{7-\delta}$, and to further demonstrate the performance and versatility of the apparatus, we also show an
example of zero-field electron-spin-resonance spectroscopy.

\begin{figure}[ht]
\includegraphics*[width=1.6in]{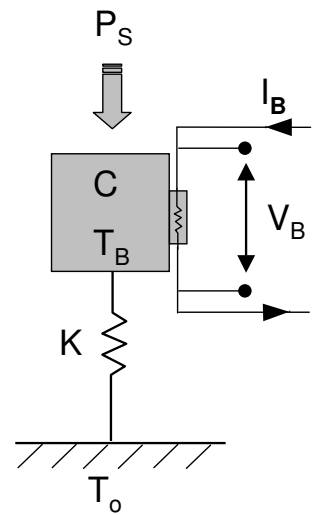}
\caption{\label{fig:model} A simple thermal model consisting of a heat capacity $C$
thermally isolated from base temperature $T_0$ by a weak thermal link of
conductance $K$. The resistive bolometer is thermally anchored to $C$ and
monitors its temperature, $T_B$ , which is elevated above $T_0$ by a constant current bias
$I_B$ passing through the bolometer. The absorption of incident signal power
$P_s$ causes heating in $C$, detectable as a temperature rise through a change in the voltage $V_B$.}
\end{figure}

\section{Bolometric Detection}

 It is instructive to calculate the minimum power detectable by a
simple thermal stage, the temperature of which is monitored by a resistive
bolometer, as depicted in Fig.~\ref{fig:model}.\cite{footnote} The
bolometer has a resistance $R_B$ and is in thermal equilibrium with a larger
heat capacity $C$ representing contributions from the sample, its holder, and
the weak thermal link. This combination is weakly connected, via a thermal
conductance $K$, to a heat sink maintained at base temperature $T_0$. The
bolometer is heated to its operating temperature $T_B$ by a bias power
$P_B=I_B^2 R_B$, where $I_B$ is the fixed bolometer bias current.  For this analysis we do not consider feedback effects,
although they are very important in the special case of transition edge
bolometers.\cite{Gildemeister} As a result, we consider a configuration
where $I_B$ provides only modest self-heating of the
bolometer, such that $\gamma \equiv (T_B-T_0) / T_0 \lesssim1$. An incident
signal power $P_S$ raises the temperature by an amount $\delta T_B = P_S/K$,
causing a change in the readout voltage across the bolometer $\delta V_B =
I_B(\textrm{d}R_B / \textrm{d}T)\delta T_B = I_B(\textrm{d}R_B /
\textrm{d}T)P_S / K$. We then define a threshold detectable signal level
that is equal to the thermal noise $\upsilon_n$ generated in a bandwidth $\Delta \nu$ in the bolometer, $\overline{\upsilon_n^2}=4k_B T_B R_B \Delta \nu$. It is then possible to write an expression for the minimum
detectable power $P_S^{min}$ in terms of the dimensionless sensitivity of the
bolometer $S_d=T/R \left|\textrm{d}R/\textrm{d}T \right|$, typically of the order of unity,
the noise power
$P_n=k_BT_B\Delta\nu$, and the bolometer bias power $P_B$:
\begin{equation}\label{eqn:minpower}
P_S^{min}={2 \over \gamma S_d} \sqrt{P_nP_B}.
\end{equation}
From this expression one immediately sees that it is desirable to minimize both
the bias and noise powers, within the combined constraints of maintaining the bolometer
temperature at $T_B$ and keeping the thermal response time fixed at a suitably
short value. By miniaturization of the sample holder, the bias power required
to reach a given temperature can be considerably reduced, while at the same
time maintaining a practical thermal time constant. The noise power is limited
intrinsically by the thermal (Johnson) noise from the bolometer resistance at temperature $T_B$.  However, most real sensors show substantial excess noise, and the Cernox 1050 sensor\cite{cernox} used in the present implementation of our experiment is no exception, showing approximately 40~dB of excess noise in the presence of a 1.3~$\mu$A bias current.  This completely accounts for the discrepancy between the minimum detectable power at 1.3~K of 17~fW calculated using Eq.~\ref{eqn:minpower} assuming only Johnson noise, and the experimentally determined value of 1.5~pW.

\section{Broadband Absorption Measurement Apparatus}

\begin{figure}[ht]
\includegraphics[width=\columnwidth]{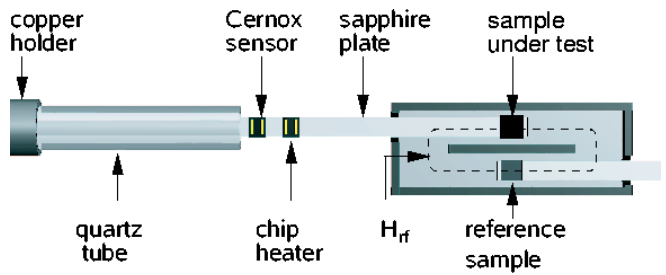}
\caption{\label{fig:endwall} Schematic cross-section of the terminated coaxial
line region showing the sample and reference materials suspended on sapphire
plates in symmetric locations in the rf magnetic field. The sapphire plate is
epoxied into the bore of a quartz tube which thermally isolates it from the copper holder,
fixed at the temperature of the 1.2~K helium bath.}
\end{figure}

 For our method of bolometric detection to be most useful, it is necessary
to deliver microwaves to the sample across a broad range of frequency and, at
the same time, not only accurately control the polarization of the microwave
field at the sample, but also maintain a fixed relationship between the field
intensity at the sample under test and the field intensity at the reference
sample. Essential to this is the design of the microwave waveguide.  We use a
custom-made transmission line, shown in cross-section in
Fig.~\ref{fig:endwall}, that consists of a rectangular outer conductor that measures 8.90~mm $\times$ 4.06~mm in cross-section and a
broad, flat centre conductor, or septum, that measures 4.95~mm $\times$ 0.91~mm. This supports a TEM mode in which the
magnetic fields lie in the transverse plane and form closed loops around the centre conductor, setting a fixed relationship between the microwave field strengths on either side of the septum. The line is terminated by shorting the centre conductor and outer conductor with a flat,
metallic endwall. This enforces an electric field node at the end of the
waveguide, adjacent to which we locate the small platelet sample and reference, with
their flat faces parallel and very close to the endwall. The broad centre
conductor ensures spatially uniform fields over the dimensions of the sample,
making it possible to drive screening currents selectively along a chosen
crystallographic direction.  The electrodynamics of the rectangular waveguide are discussed in more detail in Appendix~\ref{app:modes}. A strong variation in the power delivered
to the sample as a function of frequency arises due to standing
waves in the microwave circuit. In order to
properly account for this, we have incorporated an \textit{in-situ} normal-metal reference sample of known surface resistance that acts as an absolute
power meter. This second sample is held in a position that is
electromagnetically equivalent to the that of the test sample, on a separate
thermal stage.

\begin{figure}[ht]
\includegraphics*[width=\columnwidth]{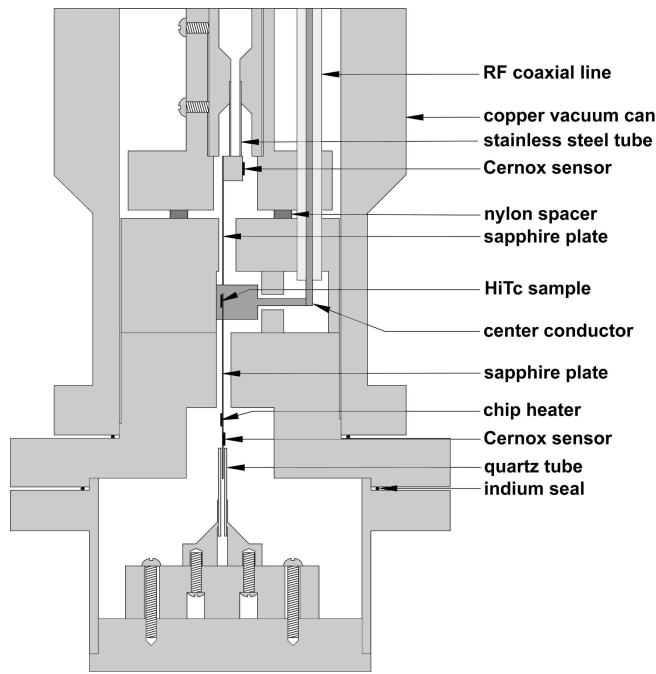}
\caption{\label{fig:fullprobe} Scale drawing of the assembled apparatus
indicating the details of the vacuum can and sample region. The alloy reference
sample is not visible in this cut-away view.}
\end{figure}

One of the challenges of cryogenic microwave absorption measurements on small,
low-loss samples is the design of the sample holder, which must measure and
regulate the sample temperature, and yet contribute negligible dissipation
compared to the sample. A widely used technique that satisfies these
requirements is that of a sapphire hot-finger in vacuum,\cite{hotfinger} allowing the
thermometry to be electromagnetically shielded from the microwave fields. In our apparatus, the sample holder is inserted through a hole that is beyond cut off for all operating frequencies. For ac calorimetric
measurements, the design of the thermal stage is critical in setting the
sensitivity of the system. The experimental arrangement is shown schematically
in Fig.~\ref{fig:endwall} with the sample under test fixed on the
end of a 100~$\mu\textrm{m}$ thick sapphire plate using a tiny amount of vacuum
grease.\cite{vacuumgrease} The plate extends 17~mm from the sample to where it is epoxied into the bore of a 1.2~mm diameter quartz glass tube that acts as a thermal weak-link to
the liquid helium bath. A Cernox thermometer and a 1500~$\Omega$ surface-mount resistor used as a heater\cite{heater}
are glued directly onto the sapphire plate with a very thin layer of Stycast
1266 epoxy,\cite{epoxy} ensuring intimate thermal contact with the sapphire and hence the
sample. We use 0.05~mm diameter NbTi superconducting electrical leads to the
thermometer and heater for their very low thermal conductance, which is in
parallel with the quartz weak-link.

The microwave circuit is powered by a Hewlett-Packard 83630A synthesized
sweeper (0.01-26.5~GHz) combined with either an 8347A (0.01-3~GHz) or 8349B
(2-20~GHz) amplifier, generating up to 23~dBm of rf power across the spectrum.
Approximately 2~m of 0.141$^{\shortparallel}$ stainless steel coaxial line\cite{coax}
delivers power from the amplifier down the cryostat to the vacuum can where it
is soldered into the rectangular line (see Fig.~\ref{fig:fullprobe}). The r.m.s. microwave magnetic field amplitude at the samples is
typically $\sim$10$^{-2}$ Oersteds, which generates $\sim\mu$K
modulations in the sample-stage temperature for a typical high quality 1~mm$^2$ high-T$_c$ sample having a low frequency $R_s$ value of 1~$\mu\Omega$.

An assembled view of the low temperature apparatus including the microwave transmission line and the positions of the sample and reference holders is provided in
Fig.~\ref{fig:fullprobe}. The sapphire plates that support both the test sample and
reference sample are inserted through 4~mm cut-off holes into the microwave
magnetic field.  The rectangular coaxial line consists of a centre conductor
made from a 0.91~mm thick copper plate soldered at one end onto the centre
conductor of the 0.141$^{\shortparallel}$ semi-rigid coaxial line, and at the
other end into the wall of the copper cavity that comprises the outer conductor
of the transmission line. To minimize the rf power dissipated in the low
temperature section of transmission line, the entire surface exposed to
microwave radiation, including the final 15~cm of semi-rigid coaxial line, was
coated with PbSn solder, which is superconducting below 7~K. During experiments, the vacuum can is completely immersed in a pumped liquid helium bath having a base temperature of 1.2~K.

The selection of a reference material for low-frequency work must be made
carefully. Initially, we chose samples cut from commercially available
stainless-steel shim stock, a common choice in infrared spectroscopy work.
Calibration experiments produced erratic results which were eventually traced
to the presence of anisotropic residual magnetism in the stainless steel.
Subsequently, we produced our own reference material, choosing an Ag:Au alloy
(70:30 at.\% made from 99.99\% pure starting materials), because it exhibits
a very simple phase diagram that guarantees homogeneity.\cite{alloy}  By using
an alloy, we ensure that the electrodynamics remain local at microwave
frequencies, avoiding the potential complications arising from the anomalous
skin effect.\cite{abrikosov} Our sample was cut from a 93$\pm5$~$\mu$m thick
foil having a measured residual dc resistivity value of $\rho$=5.28$\pm0.3~\mu\Omega$cm,
constant below 20~K.

While the thermal stage for the reference sample is similar in design to that used for the sample under test, it uses a higher conductance stainless steel thermal weak-link (in
place of the quartz tube), since the dissipation of the normal metal calibration
sample is orders of magnitude larger than that of a typical superconducting sample.
Because the apparatus was implemented as a retro-fit to an existing
experiment, the reference thermal stage had to be mounted directly onto the body of
the transmission line structure. Although the cavity walls are superconducting to reduce their absorption, we use a nylon spacer to thermally isolate the base of the reference from the transmission line to avoid direct heating. The heat-sinking of the reference base to the helium bath is made using a separate copper braid that is not visible in Fig.~\ref{fig:fullprobe}.

 As considered previously in our generic analysis, we operate the Cernox
bolometer with a constant dc current bias, typically a few $\mu$A, provided by
the series combination of an alkaline battery (1.5~V or 9~V) and bias resistor
whose value is much larger than that of the Cernox sensor. With the helium bath
under temperature regulation, the choice of bias power sets the temperature of
the sample for a given experiment, with no other temperature control necessary. All
electrical leads into the cryostat are shielded twisted pairs of insulated
manganin wire, and true four-point resistance measurements are made on all
sensors.  The voltage signal appearing on the Cernox thermometer is amplified outside the
cryostat by a carefully shielded and battery-powered circuit.  We use a
two-stage cascaded amplifier with one Analog Devices AD548 operational
amplifier per stage, chosen because these are readily available, low-noise amplifiers. The dc level is nulled between stages to prevent saturation,
and the total gain is 10$^4$. The amplified signal, corresponding
to the temperature modulation of the sample, is then demodulated with a Stanford
Research Systems SR850 digital lock-in amplifier that is phase-locked with the
rf-power amplitude modulation. There are two such systems, one for the sample
and one for the reference measurements.  The entire experiment is operated under
computer control when collecting data.

\section{Calibration}

\begin{figure}[ht]
\includegraphics*[width=0.9\columnwidth]{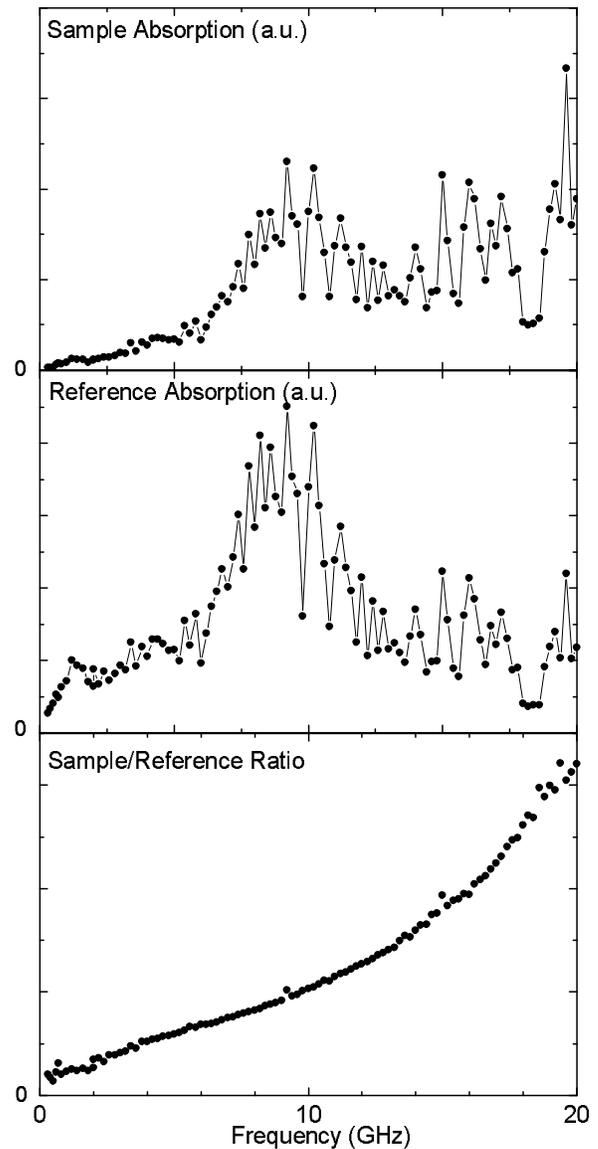}
\caption{\label{fig:ratioplot} Raw absorption spectra corresponding to the
temperature rise of the sample. Taking the ratio of the two signals accounts
for the strong frequency dependence of $H_{rf}$ introduced by standing waves in
the transmission line. The remaining frequency dependence of the ratio is due
to the different $R_s(\omega)$ spectra of the two samples.}
\end{figure}

\begin{figure}[ht]
\includegraphics*[width=\columnwidth]{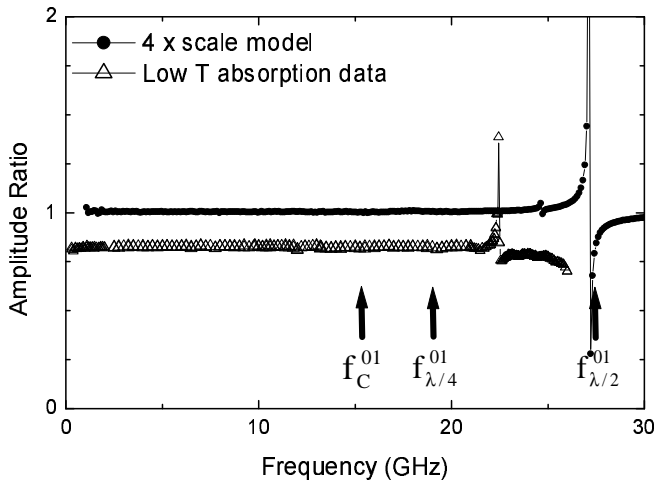}
\caption{\label{fig:metalcal} Ratio of the sample absorption to reference absorption
for identical samples, compared to measurements of the field
amplitude at equivalent positions in a $4 \times$ scale model. (Frequencies for the scale model have been scaled by a factor of 4 for the comparison.) The ratio technique is seen to break down
with a sharp resonance in both cases.  The origin of these resonances, which limit the useful frequency range of the apparatus, is discussed in detail in Appendix~\ref{app:modes} and shown to be due to the presence of standing waves of the TE$_{01}$ waveguide mode. For this mode, the three arrows indicate: the cut-off frequency $f_C^{01} = 15.38$~GHz at which the mode is first free to propagate; its quarter-wave resonance frequency $f_{\lambda/4}^{01} = 19.11$~GHz, for open-circuit termination conditions; and the half-wave resonance frequency $f_{\lambda/2}^{01} = 27.4$~GHz, for short-circuit termination conditions.  The $\lambda/4$ and $\lambda/2$ resonance frequencies bracket the observed resonances.  The scale model, which has a large transition capacitance between circular and rectangular coax sections, is seen to fall at the high end of the range.}
\end{figure}

Two steps are necessary for an absolute calibration of the surface resistance
of an unknown specimen from the measured temperature-rise data. The first is
to calibrate the absolute power sensitivity of the sample and reference thermal
stages at the actual operating temperature and modulation frequency. This is
achieved using the small \textit{in-situ} heater to drive well-characterized
heat pulses that mimic absorption by the sample, while at the same time
measuring the corresponding temperature response. The second step requires the
calibration of the magnetic field strength at the sample, at each frequency,
using the known absorption of the reference sample.  We exploit the fact that
the metallic reference sample experiences the same incident microwave field
$H_{rf}$ as the sample under test, guaranteed by conservation of magnetic
flux. This ensures that taking the ratio of the absorbed power per unit
surface area of each sample provides the ratio of the surface resistance
values:
\begin{equation}
\ {P_{abs}^{sam} \over P_{abs}^{ref}} = {R_{s}^{sam} A^{sam} \over R_{s}^{ref} A^{ref}}.
\label{eqn:ratio}
\end{equation}
The surface resistance of the unknown sample $R_{s}^{sam}(\omega)$ is then
trivially found by multiplying the power-absorption ratio, shown in
Fig.~\ref{fig:ratioplot}, by $R_s(\omega)$ of the metallic reference sample calculated
using the classical skin-effect formula $R_s(\omega)=\sqrt{\omega\mu_o \rho/
2}$ where $\omega/2\pi$ is the frequency and $\mu_o$ is the permeability of free space. The raw power-absorption spectra, shown in the first two panels of
Fig.~\ref{fig:ratioplot}, highlight the necessity of the reference sample. The
absorption spectra of the samples is completely masked by the large amplitude
variations of $H_{rf}$ caused by the standing waves in the microwave circuit.

\begin{figure}[ht]
\includegraphics*[width=\columnwidth]{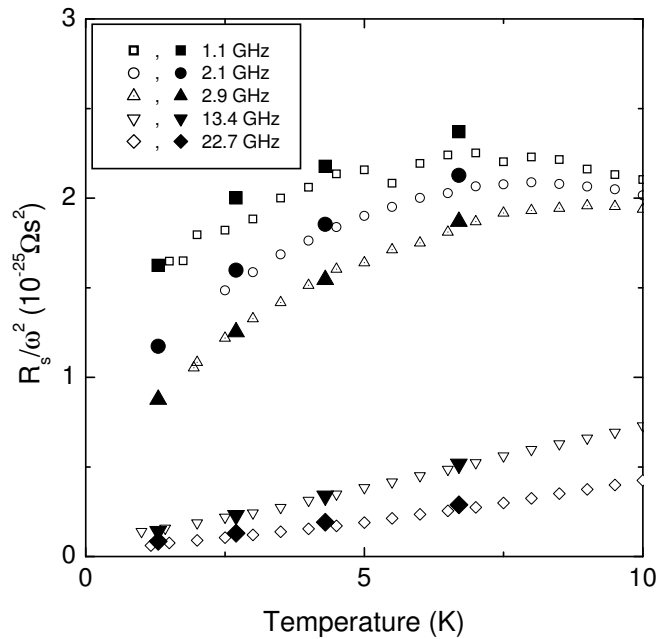}
\caption{\label{fig:resonators} Comparison of measurements made on the same
sample of YBa$_2$Cu$_3$O$_{6.50}$ using the broadband bolometric experiment (solid symbols)
with those from five microwave resonators (open symbols).  The agreement between methods is excellent. The data is plotted as
$R_s(\omega)/\omega^2$ to remove the frequency dependence associated with
superfluid screening. }
\end{figure}
An essential test of the method is to make a frequency-scanned
measurement with identical samples mounted on the sample and the reference
stages. The result should be a frequency-independent ratio across the spectrum,
equal to unity for samples with the same surface area. In
Fig.~\ref{fig:metalcal} we show such a measurement using two thin-platelet
samples of our Ag:Au reference alloy, both at the base temperature of 1.2~K.
The data reveal a ratio of 0.82 in the cryogenic apparatus, due to the fact
that the centre conductor is offset from centre by 0.1~mm in the termination
region, intensifying the fields on one side relative to the other. This scale
factor must be included in the calibration of all experimental data. The sharp
resonance seen in the ratio at 22.5~GHz indicates the presence of a non-TEM
electromagnetic mode in the sample cavity that breaks the symmetry in field strength between sample and reference positions. For our present design, this sets the
upper frequency limit of operation.

In an attempt to gain further insight into the field configurations in the
transmission line, and to understand how the higher order waveguide modes limit
the upper frequency range, we built a scale model of the setup having all
dimensions larger than those of the cryogenic apparatus by a factor of four. For
comparison, a frequency scan of the model structure is included in
Fig.~\ref{fig:metalcal}, using loop-probes in the positions of the samples. The
data show that the non-TEM-mode resonance occurs at 27~GHz, considerably higher
than in the low temperature experiment. It turns out that the breakdown of the
sample-reference symmetry occurs not at the frequency at which higher order
waveguide modes first propagate in our structure, but at the frequency at which
they form a resonant standing wave. A full discussion of this is given in
Appendix~\ref{app:modes}.

A number of other experimental tests were important to verify the proper
operation of the system. Frequency scans without samples mounted on the
sapphire stages confirmed that background absorption due to the sapphire
and tiny amount of vacuum grease used to affix the samples is negligible --- it is unmeasurable at low frequency, and contributes no more than 2~$\mu\Omega$ to an $R_s$ measurement at 21~GHz. Scans without a sample also confirmed that no significant leakage heat current propagates to the
thermometers directly from the microwave waveguide. The high thermal stability of the
cryostat system is due in part to the very large effective heat capacity of the
pumped 4~litre liquid-helium bath at 1.2~K. In addition, it is always important
to make certain the temperature modulations of the samples are sufficiently
small that the response of the thermal stages remains in the linear regime.
Furthermore, measurements with the same sample located in different positions
along the sapphire plate, with up to 0.5~mm displacement from the central
location in the waveguide, confirmed that there is enough field homogeneity
that our sample alignment procedure using an optical microscope is sufficient,
and that samples of different sizes experience the same fields.

A very convincing verification of the technique is provided by the ability to
compare broadband $R_s(\omega,T)$ data with measurements of the \textit{same
sample} in five different high-Q microwave resonators. These experiments probe
the temperature dependence of the absorption to high precision at a fixed
microwave frequency: however, the determination of the {\em absolute} value of $R_s$ is limited to about 10\% as discussed previously. The bolometric method has the advantage of being able to measure a true spectrum because the dominant uncertainty, the absolute surface resistance of the reference sample, enters as a scale factor that applies across the entire frequency and temperature range. A detailed discussion of the uncertainties in the bolometry data will be presented in the subsequent section. Figure~\ref{fig:resonators} shows that there is very good agreement of both the temperature and frequency dependence of the surface resistance as measured independently by the fixed-frequency and broadband experiments.

\begin{figure}[ht]
\includegraphics*[width=\columnwidth,height=.95\columnwidth]{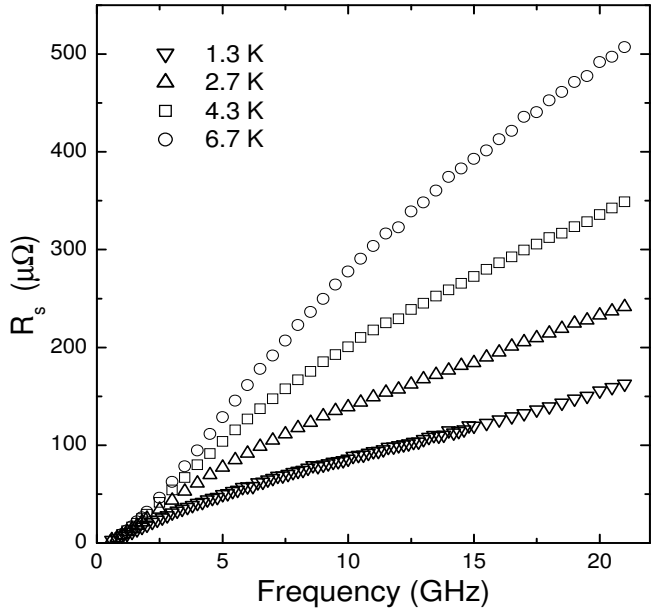}
\caption{\label{fig:Rsplot} Broadband measurements of the microwave surface
resistance spectrum of YBa$_2$Cu$_3$O$_{6.50}$ obtained with the bolometric
apparatus below 10~K. The low frequency absorption approaches the resolution
limit of the apparatus, while the upper frequency limit is imposed by the
resonance in the microwave structure.}
\end{figure}
\begin{figure}[t]
\includegraphics*[width=.95\columnwidth]{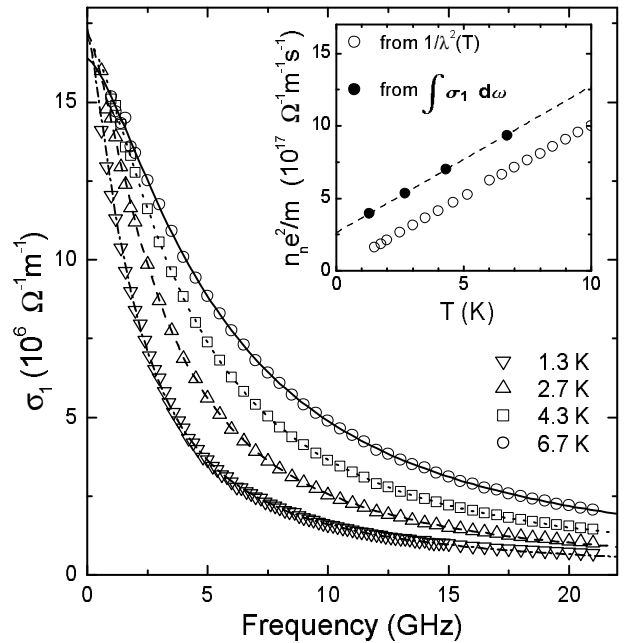}
\caption{\label{fig:Sigma1} The real part of the microwave conductivity
$\sigma_1(\omega,T)$ extracted from the broadband $R_s(\omega,T)$ measurements.
We use the self-consistent fitting procedure described in Appendix~\ref{app:kramers} to properly account for contributions to field-screening by the quasiparticles.}
\end{figure}

\section{Performance}

Figure~\ref{fig:Rsplot} presents an example of high resolution broadband
measurements of the frequency-dependent and temperature-dependent surface resistance of a
superconducting sample.\cite{turner} This particular data set is for $\hat{a}$-axis currents
in a YBa$_2$Cu$_3$O$_{6.50}$ single crystal (T$_c$=56~K) having dimensions
1.25$\times$0.96$\times$0.010~mm$^3$. The data span the range 0.6-21~GHz,
limited at high frequency by the resonance in the system, and at low frequency
by the small dissipation of the sample, which approaches the resolution limit of
the experiment. At 1~GHz, the values for the statistical r.m.s. uncertainty in surface resistance, $\delta R_s$, are about 0.2, 0.4, 0.6, and 1.3~$\mu\Omega$ for $T$~=~1.3,
2.7, 4.3, and 6.7~K respectively. Error bars have been omitted from the figure
for clarity. Systematic contributions to the uncertainty enter as overall scale
factors in the $R_s$ data and are attributed to an uncertainty in the DC
resistivity of the thin Ag:Au alloy foil used as a reference sample
($\sim$5\%), the surface area of the samples ($\sim$1\%), and the absolute
power sensitivity of the thermal stage ($\sim$1\%).

\begin{figure}[ht]
\includegraphics*[width=.95\columnwidth]{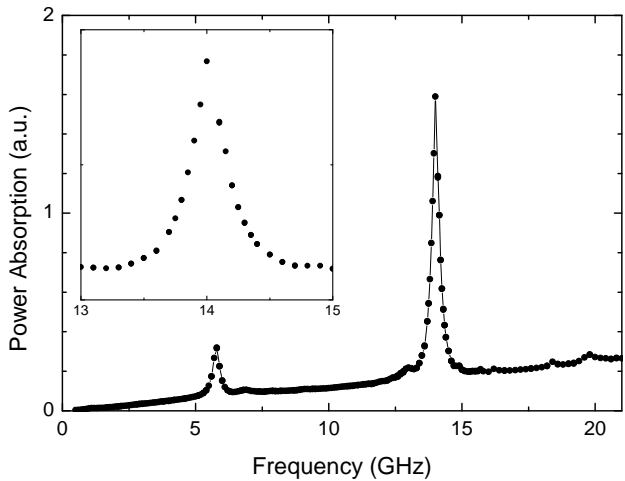}
\caption{\label{fig:esr} The electromagnetic absorption spectrum of a
Y$_{0.99}$Gd$_{0.01}$Ba$_2$Cu$_3$O$_{6.99}$ single crystal at 1.3~K. The
spectrum consists of a broad background due to the quasiparticle absorption in
the superconductor in addition to the zero field ESR lines generated by the low
concentration of magnetic Gd$^{3+}$ impurities. Only ions residing within a
distance $\lambda$ of the crystal surface contribute to the signal as
the applied rf field is strongly screened by the superconductor.}
\end{figure}
The frequency dependence observed in $R_s(\omega)$ is due to absorption by
quasiparticles thermally excited from the superfluid condensate. The quantity of fundamental theoretical interest is the real part of the conductivity spectrum
$\sigma_1(\omega)$, which must be extracted from the experimentally measured
$R_s(\omega)$ data. A thorough discussion of the method we use to do this is
given in Appendix~\ref{app:kramers}  but, to first approximation, the shape of the conductivity spectrum
can be found by dividing $R_s(\omega)$ by a factor of $\omega^2$ to account for
the screening of the applied field by the superfluid. Figure~\ref{fig:Sigma1}
shows the conductivity spectra extracted from the $R_s(\omega)$ data using the
complete analysis. It is immediately apparent why improving the sensitivity
of the experiment is of the utmost importance. The low frequency region, where
the power absorption becomes very small, is where the conductivity exhibits the
strongest frequency dependence and is most important to measure accurately. The
spectrum at 1.3~K has a width of the order of 5~GHz, signifying very long quasiparticle
scattering times, indicative of the high quality of our
YBa$_2$Cu$_3$O$_{6.50}$ crystal. For this very clean sample, most of the
spectral weight resides below the experimental frequency limit of 21~GHz at
1.3~K, but the increase in scattering with increasing temperature quickly
broadens the spectra thus motivating future designs capable of probing a
broader frequency range. Many cuprate materials, such as
Bi$_2$Sr$_2$CaCu$_2$O$_{8+\delta}$, have scattering rates that are orders of
magnitude higher and require THz frequency techniques to probe their
dynamics.\cite{corson}

As a final demonstration of the sensitivity of the broadband
instrument we have described, we include a frequency scan of a superconducting sample that
exhibits clearly discernable absorption lines due to zero field electron spin
resonance (ESR) of a low density of magnetic impurities. Gadolinium ions
(electron spin $S=7/2$) substitute for yttrium, sandwiched between the two CuO
planes in the YBaCuO unit cell, and the splitting of the degenerate Gd$^{3+}$
hyperfine levels by the crystalline field provides a very sensitive probe
of the local microscopic structure.\cite{janossy} These measurements are
typically performed in a high field spectrometer, but the bolometric system
provides a means of performing zero-field measurements. Figure~\ref{fig:esr}
shows the 1.3~K absorption spectrum of a 1~mm$^2$
Y$_{0.99}$Gd$_{0.01}$Ba$_2$Cu$_3$O$_{6.99}$ single crystal consisting of a
broad background due to the quasiparticle conductivity, essentially unaltered
by the presence of the Gd ions, with the ESR absorption peaks superposed. The
high signal-to-noise ratio achieved with the experiment allows one to resolve the
spectrum in great detail.

The apparatus described here has sufficient sensitivity and frequency range for it to be immediately applicable to many other interesting problems in condensed matter physics.  These might include: the study of low-lying collective modes in metals and superconductors; zero-field electron spin resonance in correlated insulators; and the study of critical phenomena at the metal--insulator transition and near the zero-temperature magnetic critical points that occur in certain $d$- and $f$-electron metals.  With a little attention to thermal design, specifically the thermal separation of sample stages from the microwave waveguide, the superconducting coatings on the waveguide could be removed and the system used in high magnetic fields.  This would open interesting possibilities in the area of metals physics, such as high-resolution cyclotron and periodic-orbit resonance, as well as the study of vortex dynamics and vortex-core spectroscopy in superconductors.  Finally, further miniaturization of the experiment should also be possible: the ultimate goal would be to extend the frequency range of this type of spectroscopy so that is joins seamlessly on to the THz range now accessible using pulsed-laser techniques.

\acknowledgments

The authors are indebted to Pinder Dosanjh for his technical assistance, as
well as to Ruixing Liang for the YBaCuO samples employed in these experiments.  We
also acknowledge financial support from the Natural Science and Engineering
Research Council of Canada and the Canadian Institute for Advanced Research.

\appendix

\section{Frequency Response of Distributed Thermal Stage}
\label{app:thermal}

We wish to calculate the temperature response of a simple thermal stage to a
sinusoidal heat flux $P_{in}=\textrm{Re} \left\{\tilde{P_0}e^{\textrm{\small i} \omega t}\right\}$ superimposed on a static temperature gradient. We consider an arrangement where the isothermal sample stage has neglible thermal mass and is connected to base temperature by a weak thermal link with distributed heat capacity  $c_V$ per unit volume. It is a straightforward extension to include an additional lumped heat capacity for the isothermal stage; once the lumped heat capacity dominates, the frequency response simplifies to that of a single-pole low-pass filter.  However, in our case this is unnecessary: for electrodynamic measurements at low temperatures, the sample holder is required to be both electrically insulating and highly crystalline, and will therefore have very low heat capacity.

Here we consider the one-dimensional problem of a thin bar (the quartz tube in
our apparatus) of length $\ell$ and cross-sectional area $A$, with one end fixed at a base temperature $T_0$ while the other end is heated by a heat flux due to sample power absorption. The propagation of a heat current $J_Q$
through the bar is constrained by the continuity equation $ \partial J_Q/\partial x +c_V \partial T/\partial t=0$ and the thermal conductivity $\kappa$ is defined by $J_Q=-\kappa  \partial T/\partial x$.  Together, these lead to the one-dimensional heat equation $\partial T/\partial t=\alpha \partial^2 T/\partial x^2$
where $\alpha=\kappa/c_V$ is the thermal diffusivity. Defining a complex thermal diffusion length
$\tilde{\delta}=\sqrt{\alpha/\textrm{i}\omega}$, the time-dependent part of the temperature profile can be written
\begin{equation}
\label{eqn:tempprofile1}
\Delta T(x,t)=\textrm{Re} \left\{ \tilde{T}e^{\textrm{\small i} \omega t} \sinh(x/\tilde{\delta}) \right\}
\end{equation}
where $\tilde{T}$ is fixed by the heat-flux boundary condition:  $P_{in}=- \kappa A (\partial T/\partial x)\vert_{x=\ell}$. This completely determines the frequency-dependent temperature rise of the sample stage:
\begin{equation}
\label{eqn:tempresponse}
 \Delta T(\ell,t)  = \textrm{Re}\left\{ {\tilde{P}_{0} \over \kappa A} \tilde{\delta} e^{\textrm{\small i} \omega t} \tanh(\ell / \tilde{\delta} ) \right\}
\end{equation}
In the low frequency limit, the temperature rise reverts to the usual result:
\begin{equation}
\Delta T(\ell,t)={P_{0} \ell \over \kappa A}\cos(\omega t),
\end{equation}
where, without loss of generality, we have set the phase of the input heat flux to zero.
In the high frequency limit, the thermal diffusion length becomes shorter than the weak link and the temperature rise is reduced, being given by:
\begin{equation}
\label{eqn:highfreqyresponse}
\Delta T(\ell,t)={P_{0} \vert \tilde{\delta} \vert \over \kappa A}\cos(\omega t - \pi/4)
\end{equation}
where $\vert {\tilde{\delta}} \vert=\sqrt{\alpha/\omega}$. At finite frequencies, part of the heat flux is diverted into the distributed heat capacity of the thermal link.  For a fixed input power (and hence fixed temperature {\em gradient} at the end of the thermal link) this leads to smaller temperature rises and a decreased sensitivity of the bolometric method.  Clearly, the experimental sensitivity of the bolometric method will be optimized by operating in the low-frequency limit: $\vert {\tilde{\delta}} \vert > \ell$ or $\omega < \alpha/\ell^2$.  A consideration of the thermal diffusivity and dimensions of the weak-link must therefore be part of any plan to increase the modulation frequency.

\begin{figure}[ht]
\includegraphics*[width=\columnwidth]{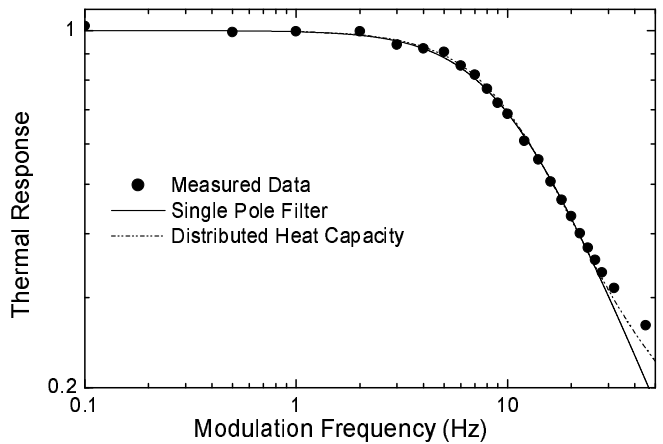}
\caption{\label{fig:ThermalResponse} Low temperature ($T = 1.3$~K) measurements of the dynamic thermal response of the quartz-tube bolometer platform.  Curves on the plot show fits using lumped and distributed heat capacity models.}
\end{figure}

Fig.~\ref{fig:ThermalResponse} shows the frequency response of the sample thermal stage in our apparatus when it was subjected to a sinusoidally varying heater power, normalized to the static response.  Included in the figure are fits to the distributed-heat-capacity model, Eq.~\ref{eqn:tempresponse} and a single-pole low-pass filter response $\Delta T(\omega)/\Delta T(\omega = 0) = 1/\sqrt{1 + \omega^2 \tau^2}$.  Although both curves fit the data well over most of the frequency range, the best-fit value of the time constant $\tau$ in the lumped-element model corresponds to a heat capacity much larger than the calculated heat capacity of the sapphire sample stage.  Instead, the value obtained from the fit is approximately half the heat capacity of the quartz tube, indicating the correct physics is that of heat diffusion in a distributed thermal system.

\section{Design Strategy for Rectangular Coaxial Transmission Line}
\label{app:modes}

\begin{figure}[ht]
\includegraphics*[width=.85\columnwidth]{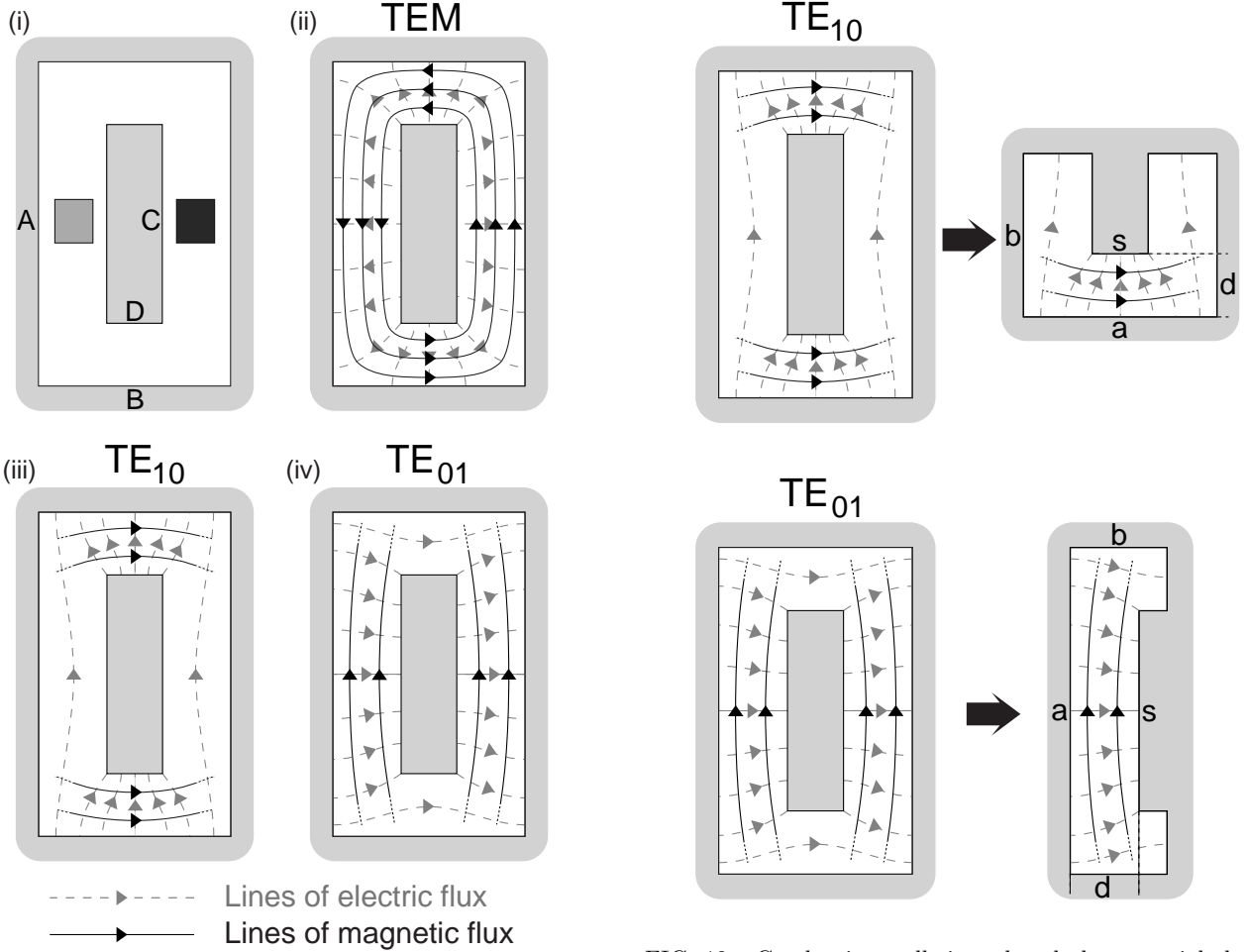}
\caption{\label{fig:CoaxModes} Cross sections of the rectangular coaxial
transmission line:  (i) physical layout of the transmission line showing
dimensions ($A,B,C,D$) of the inner and outer conductors, and sample and
reference positions (shaded squares); (ii) fields of the TEM mode, showing how
continuity of flux and a broad, flat inner conductor produce uniform, well
polarized fields of equal intensity and opposite direction at the sample and
reference positions; (iii) fields of the TE$_{10}$ mode; and (iv) fields of the
TE$_{01}$ mode. (Magnetic fields of the transverse electric modes contain a
component along the direction of propagation and do not form closed loops in
the transverse plane.)  It is clear that the TE$_{01}$ mode is most harmful to
the operation of the broadband apparatus: its magnetic fields have high
intensity at the sample and reference positions and break the balance that
otherwise exists in the TEM mode.}
\end{figure}

In optimizing a microwave transmission line for the bolometric measurement of
surface resistance the guiding aims must be: to deliver microwave power
efficiently to the sample region, over as wide a frequency range as possible
and with a well defined polarization; to have regions of uniform microwave
magnetic field at the sample and reference positions; and, at these positions,
to have a fixed, frequency-independent ratio between the field strengths.
These aims can be met by using an impedance-matched (50~$\Omega$), single-mode
coaxial line, with rectangular cross section and a broad, flat center conductor or
septum, as shown in Fig.~\ref{fig:CoaxModes}(i).  In addition, the dimensions of the rectangular coaxial line should
be chosen carefully, to prevent higher-order waveguide modes from entering the
operating frequency range of the experiment, as these modes break the symmetry
in field strength between sample and reference positions.  This appendix
outlines how to undertake the optimization.

A rectangular coaxial line, like any two-conductor line, supports a transverse
electromagnetic (TEM) wave at all frequencies.  Figure~\ref{fig:CoaxModes}(ii)
shows its electric and magnetic field configurations.  The TEM mode has the
desirable property that its magnetic fields lie in a plane perpendicular to the
direction of propagation, forming closed loops around the centre conductor.
Conservation of magnetic flux then leads to a fixed, frequency-independent
relation between the fields on either side of the septum. These fields will
also be quite homogeneous, as long as the height $C$ of the centre conductor is
large compared to the gap $(B - D)/2$ between the centre and outer conductors.
To deliver microwave power efficiently to the sample region the characteristic
impedance of the TEM mode must be close to that of the cylindrical coaxial line
used to bring microwaves into the cryostat.  Gunston \cite{gunston} has
tabulated data on the impedance of rectangular coaxial line, and gives some
useful approximate formulas. The following expression, due to Br\"ackelmann, is
stated to be accurate to 10\% for $D/B < 0.3$ and $C/A < 0.8$:
\begin{equation}
Z_0 \sqrt{\epsilon_r} = 59.952 \ln\left(\frac{A + B}{C + D}\right) \Omega,
\end{equation}
where $\epsilon_r$ is the relative permittivity of the dielectric filling the
transmission line.

\begin{figure}[ht]
\includegraphics*[width=.85\columnwidth]{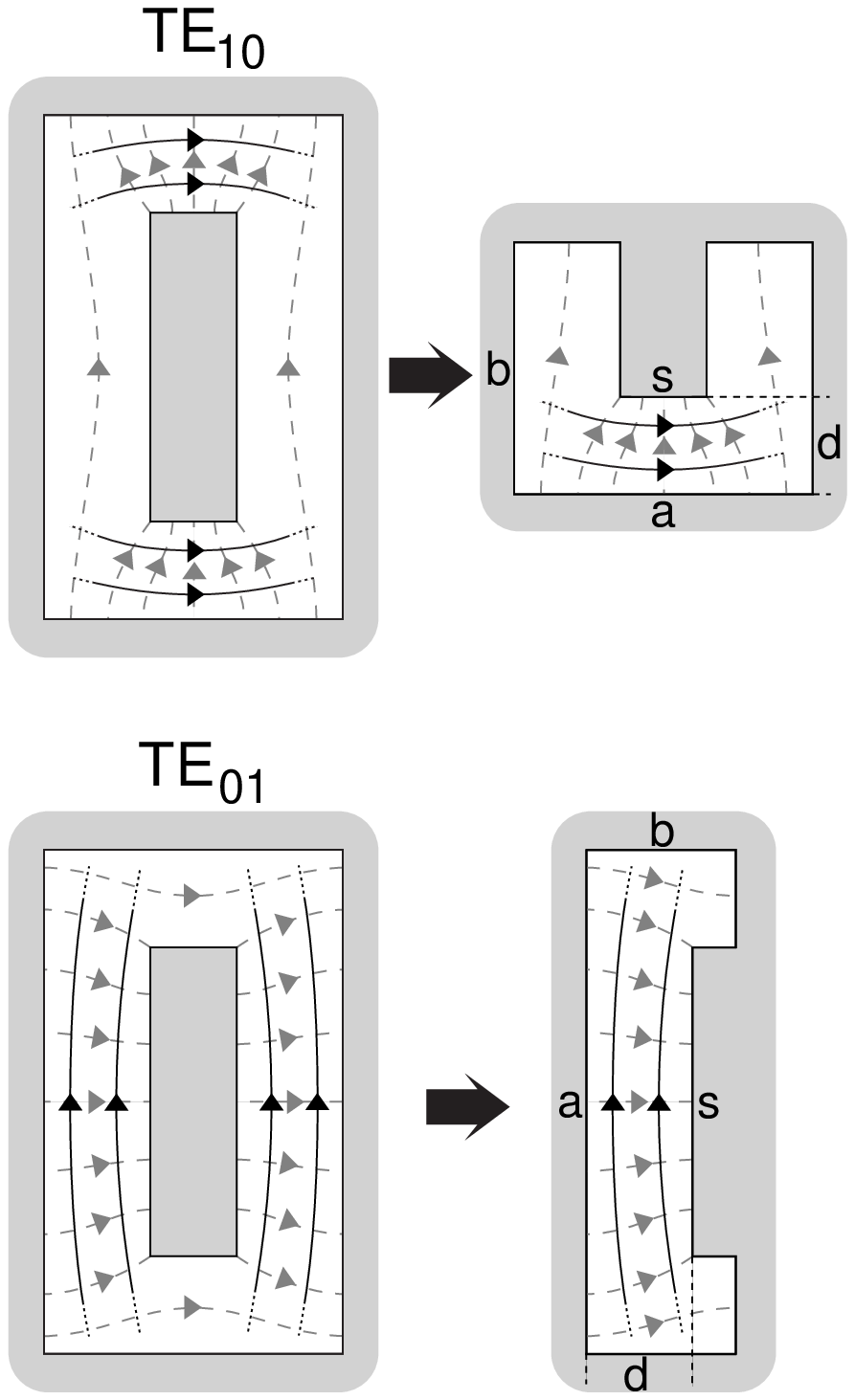}
\caption{\label{fig:CoaxModes2} Conducting walls introduced along special electric
equipotentials allow the waveguide modes of rectangular coaxial line to be
mapped onto the fundamental mode of ridged waveguide, a problem extensively
studied in the literature.  The figures show the relabelling of dimensions in
Pyle's notation~\cite{pyle} as $a,b,d$ and $s$.  }
\end{figure}

We now come to the question of what places an upper limit on the useful
frequency range of the rectangular coaxial waveguide.  At high frequencies our
method, which incorporates an in-situ power meter, suffers a spectacular
breakdown in the ratio of the relative strengths of the microwave magnetic
fields at the sample and reference positions, as shown in
Figure~\ref{fig:metalcal}.  This is caused by the presence of higher-order
waveguide modes, which have different character from that of the TEM mode under
the mid-plane reflection symmetries of the rectangular line.  The waveguide
modes with the lowest cut-off frequencies are the transverse electric modes
TE$_{10}$ and TE$_{01}$, shown in Figures~\ref{fig:CoaxModes}(iii) and
\ref{fig:CoaxModes}(iv) respectively.  These have the property that magnetic
fields on opposite sides of the septum point in the {\em same} direction.  The
fields of the TEM mode, in contrast, are {\em antiparallel}, causing an
admixture of TEM and TE modes to lack the important characteristic of equal
field intensities at sample and reference positions.  Particularly damaging is
the TE$_{01}$ mode, which is not screened by the septum and has high field
intensity in the vicinity of the sample and reference. In principle it is
possible to avoid exciting the transverse electric modes by building a very
symmetric transmission line.  In practice, however, we find this to be
impossible --- sufficiently large symmetry-breaking perturbations are always
present. Nevertheless, maintaining high symmetry is still desirable.  A
comparison of our results with calculations of the cut-off frequencies of the
transverse electric modes shows that at frequencies where the higher order
modes are free to propagate, they do not immediately cause a breakdown in field
ratio: this only occurs when the transverse electric modes come into resonance.
(This can be seen very clearly in Figure~\ref{fig:metalcal}.) As a result, the
range of operating frequency can be extended by as much as 50\% just by
shortening the final section of transmission line and carefully designing the
transition between the cylindrical and rectangular sections.

Optimizing the range of single-mode operation of the rectangular transmission
line requires a method for calculating the cut-off frequencies of the TE$_{10}$
and TE$_{01}$ modes.  While waveguide modes in two-conductor rectangular
transmission lines have not been extensively studied, their field
configurations can be mapped onto a more common geometry: that of ridged
waveguide. Figure~\ref{fig:CoaxModes2} shows how.  Electric equipotentials run
perpendicular to lines of electric flux, and special equipotentials,
corresponding to local minima of the magnetic flux density, exist on the
symmetry axes of the rectangle. A conducting wall can be introduced along these
lines without disturbing the field distributions, thereby mapping each mode
onto an equivalent ridged waveguide. Figure~\ref{fig:CoaxModes2} illustrates
the two different ways this is done, for the TE$_{10}$ and TE$_{01}$ modes
respectively.  A very early calculation of the cutoff frequency of ridged
waveguide was carried out by Pyle~\cite{pyle} and is notable for its
simplicity, generality and enduring accuracy when compared to more recent numerical
methods.\cite{amari} Pyle's approach is to solve for the transverse resonance
condition of the waveguide, which is equivalent to finding the cut-off
frequency $\omega_c$.  We have used this method in our design process, as it is
easy to implement (involving only algebraic equations) and is accurate to
several percent except when the septum becomes very thin.  The length $\ell$ of
the rectangular line, the cut-off frequency, and the discontinuity capacitance
of the cylindrical-to-rectangular transition together determine the resonant
frequencies $\omega_R$ of the transverse electric modes. There are two limiting
cases, corresponding to open-circuit ($\ell = \lambda/4$) and short-circuit
($\ell = \lambda/2$) termination (where $\lambda = 2 \pi/k$ is the wavelength
along the guide), that follow from the waveguide dispersion relation.
\begin{eqnarray}
\omega_R^2 &= &\omega_c^2 + c^2 k^2 = \omega_c^2 + \frac{4 c^2
\pi^2}{\lambda^2} \\
 & = & \omega_c^2 + \frac{c^2 \pi^2}{4 \ell^2}:~{\rm open~circuit}\\
 & = & \omega_c^2 + \frac{c^2 \pi^2}{\ell^2}:~{\rm short~circuit}
\end{eqnarray}
A high capacitance for the TE modes at the transition from cylindrical to
rectangular coax is clearly favourable: it better approximates the short
circuit termination condition and leads to resonant frequencies at the upper
end of the range. This effect is responsible for the difference in resonant
frequencies between the scale model and the  actual apparatus seen in
Figure~\ref{fig:metalcal}.  There is, however, a trade-off to be made: too
large a transition capacitance for the TEM mode will result in most of the
microwave power being reflected before it reaches the sample.
The dimensions of the rectangular guide in our apparatus are $A = 8.90$~mm, $B = 4.06$~mm, $C =4.95 $~mm, $D =0.91 $~mm and $\ell = 6.60$~mm.  The cut-off frequencies for the TE$_{10}$ and TE$_{01}$ modes are calculated to be 19.68~GHz and 15.38~GHz respectively.  The quarter wave-resonances would then occur at 22.72~GHz and 19.12~GHz, and the half wave resonances at 30.06~GHz and 27.44~GHz.

\section{Extraction of $\sigma_1(\omega)$ from $R_s(\omega)$ Measurements}
\label{app:kramers}

In this appendix we show how the microwave conductivity spectrum of a superconductor can be obtained from a measurement of its frequency-dependent surface resistance.  This process is similar to the extraction of conductivity spectra in the infra-red frequency range from reflectance measurements.  In both cases, we begin with incomplete information about the electrodynamic response: the bolometric technique described in this paper measures only the {\em resistive} part of the surface impedance; and optical techniques typically obtain the magnitude, but not the phase, of the reflectance.  However, the conductivity $\sigma \equiv \sigma_1 - {\rm i}\sigma_2$ is a causal response function, and its real and imaginary parts are related by a Kramers--Kr\"onig transform:
\begin{equation} \label{KK}
\sigma_{2}(\omega)={2\omega \over \pi} {\mathcal P} \int_0^\infty
{\sigma_1(\Omega) \over \Omega^2 - \omega^2} {\rm d}\Omega,
\end{equation}
where $ {\mathcal P}$ denotes the principle part of the integral.  At first sight we seem to have replaced one uncertainty, incomplete knowledge of the phase, by another, the finite frequency range over which the measurements have been made.  However, a suitable extrapolation of the data out of the measured frequency range is usually possible and makes the transform a well-defined procedure in practice.

 We consider the limit of local electrodynamics, in which the microwave surface impedance $Z_s$ is
related to the complex conductivity in a straightforward manner by the
expression
\begin{equation} \label{eqn:Zs}
Z_s=R_s+\textrm{i}X_s=\sqrt{ {\textrm{i} \omega \mu_0} \over
{\sigma_1-\textrm{i}\sigma_2} }. \label{eqn:surfimped}
\end{equation}
Very generally, the conductivity can be partitioned into a superfluid part
$\sigma_S$, consisting of a zero-frequency delta function and an associated
reactive term, and a normal-fluid component $\sigma_N$:
\begin{eqnarray}
\sigma(\omega,T)&=&\sigma_{1S}-\textrm{i}\sigma_{2S}+\sigma_{1N}-\textrm{i}\sigma_{2N}\\
\nonumber &=&\pi  {n_s e^2 \over m^\ast}\delta(\omega)-\textrm{i}{n_se^2 \over
m^\ast \omega} + \sigma_{1N}-\textrm{i} \sigma_{2N},
\end{eqnarray}
where $m^\ast$ is the quasiparticle effective mass.  In the clean-limit, where
the quasiparticle scattering rate $1/\tau$ is much less than the spectroscopic
gap $2 \Delta$, sum-rule arguments enable a clean partitioning of the
conduction electron density $n$  into a superfluid density  $n_s$ and a
normal-fluid density $n_n = n - n_s$.  In this type of generalized two-fluid
model,\cite{berlinsky} the temperature dependence of $n_s$ is determined
phenomenologically from measurements of the magnetic penetration depth
$\lambda$ through the relation
\begin{equation}
n_s(T)e^2/m^\ast=[\mu_0\lambda^2(T)]^{-1}.
\end{equation}
Applying this, we can write the conductivity at finite frequencies as
\begin{equation} \label{eqn:sigma}
\sigma(\omega,T)=\sigma_{1N}(\omega,T)-\textrm{i}\left[\sigma_{2N}(\omega,T)+{1
\over \mu_0\omega\lambda^2(T)}\right].
\end{equation}
From Eq.~\ref{eqn:surfimped} it is clear that $R_s(\omega)$ is determined by
both the real and imaginary parts of the conductivity. However, one
simplification occurs at temperatures well below $T_c$, where few thermally
excited quasiparticles exist, and the low frequency reactive response is
dominated by the superfluid. In this case, a good approximation to the
relations becomes
\begin{eqnarray}\label{RX}
R_s(\omega,T)&=&{1 \over 2}\mu_0^2\omega^2\lambda^3(T)\sigma_1(\omega,T),\\
\nonumber X_s(\omega,T)&=&\mu_0\omega\lambda(T).
\end{eqnarray}
At higher frequencies and temperatures, a more complete treatment would account
for quasiparticle contributions to field screening, which enter through
$\sigma_{2N}(\omega,T)$. We use an iterative procedure to obtain the quasiparticle conductivity spectrum $\sigma_{1N}(\omega)$, starting from the good initial guess provided by Eq.~\ref{RX}.  The process goes as follows.  A phenomenological form that captures the key characteristics of the dataset but has no physical motivation,
namely $\sigma_1(\omega)=\sigma_0/[1+(\omega/\Gamma)^y]$, is fitted to the spectrum and used to extrapolate out of the measured frequency range.  The Kramers--Kr\"onig transform (Eq.~\ref{KK})  can then be applied to obtain $\sigma_{2N}(\omega)$.  With $\sigma_{2N}(\omega)$ in hand, and with the superfluid contribution to $\sigma_2$ known from measurements of the magnetic penetration depth, a new extraction of the conductivity from the $R_s(\omega)$ data is made, this time using the {\em exact} expression, Eq.~\ref{eqn:surfimped}.  The whole procedure is repeated to self-consistency.  We find that the procedure is stable and converges rapidly, and is not sensitive to the details of the high-frequency extrapolation.  Also, the corrections are quite small for the low
temperature dataset shown in Fig.~\ref{fig:Sigma1}: at the highest temperature and frequency they amount to a 7\% change in $\sigma_1$.

In addition, two independent experimental checks give us further assurance that we obtain the correct conductivity spectra.  We first note that Eq.~\ref{eqn:surfimped} contains an expression for the surface reactance $X_s$.  Therefore, once we have measured the penetration depth and extracted the conductivity spectra from the $R_s(\omega)$ data, we can predict the temperature dependence of the surface reactance at {\em any} frequency and compare with experiment.  We have made this comparison at 22.7~GHz, with surface reactance data obtained on the same crystal, and find excellent agreement.  We note that this is a frequency high enough for quasiparticle scattering to have a discernible effect on the surface reactance.

A second verification of the conductivity extraction procedure is its ability to predict the spectral weight that resides \textit{outside} the frequency window of the measurement.   A corollary of the Kramers--Kr\"onig relation~\ref{KK} is the oscillator-strength sum rule
\begin{equation}\label{eqn:sumrule}
{n e^2 \over m^\ast}= {2 \over \pi} \int_0^\infty \sigma_1(\omega,T) {\rm d}\omega.
\end{equation}
In a superconductor, in the clean limit, the sum rule requires that any spectral weight disappearing from the superfluid density $n_s(T)$ as temperature is raised must reappear as an increase in the frequency-integrated quasiparticle conductivity.  We have carried out this comparison, which is shown in the inset of Fig.~\ref{fig:Sigma1}.\cite{turner} The good agreement in the temperature dependence of the superfluid and normal-fluid densities is a strong verification of both the analysis procedure {\em and} the bolometric technique.


\begin{thebibliography}{28}
\bibitem{coax} We use standard 50~$\Omega$ impedance semi-rigid coaxial line having a Teflon dielectric and a 0.141$^\shortparallel$ diameter outer conductor made from stainless steel and 0.0359$^\shortparallel$ diameter inner conductor made from silver-plated, copper-clad steel.
\end{thebibliography}
\end{document}